# Quantifying photometric observing conditions on Paranal using an IR camera

Florian Kerber*[a], Richard R. Querel[b,c], Reinhard Hanuschik[a]

[a]European Southern Observatory, Karl-Schwarzschild-Str. 2, 85748 Garching, Germany;
[b]National Institute of Water and Atmospheric Research (NIWA), Lauder, New Zealand;
[c]Department of Electrical Engineering, University of Chile, Santiago de Chile, Chile

Abstract

A Low Humidity and Temperature Profiling (LHATPRO) microwave radiometer, manufactured by Radiometer Physics GmbH (RPG), is used to monitor sky conditions over ESO's Paranal observatory in support of VLT science operations. In addition to measuring precipitable water vapour (PWV) the instrument also contains an IR camera measuring sky brightness temperature at 10.5 µm. Due to its extended operating range down to -100 °C it is capable of detecting very cold and very thin, even sub-visual, cirrus clouds. We present a set of instrument flux calibration values as compared with a detrended fluctuation analysis (DFA) of the IR camera zenith-looking sky brightness data measured above Paranal taken over the past two years. We show that it is possible to quantify photometric observing conditions and that the method is highly sensitive to the presence of even very thin clouds but robust against variations of sky brightness caused by effects other than clouds such as variations of precipitable water vapour. Hence it can be used to determine photometric conditions for science operations. About 60 % of nights are free of clouds on Paranal. More work will be required to classify the clouds using this technique. For the future this approach might become part of VLT science operations for evaluating nightly sky conditions.

## 1. INTRODUCTION

Photometry, the quantitative measurement of the brightness of a celestial object, has always been a fundamental aspect of astronomical research. The presence of clouds, of course, is a major complication in this context and observatory sites are selected – amonh other things – based on the percentage of clear and photometric nights available per year. For the astronomer it is more important to reliably know whether the sky was free of clouds during the night of observations. Hence it is a primary interest of any observatory to have real-time information on the atmospheric conditions, including the presence of clouds.

ESO's Very Large Telescope (VLT) is located on Cerro Paranal (24.6°S, 70.4°W, 2635 m.asl). The observatory has been operating instruments to monitor ambient conditions for many years, see http://archive.eso.org/asm/ambient-server. Historical data are available through the ESO archive. The latest addition to this suite of instruments is a water vapour monitor in support of VLT science operations[1]. Atmospheric water vapour content can be expressed as precipitable water vapour (PWV) which is the equivalent condensed amount of water in an atmospheric column above the observer (as a depth in units of mm). Paranal is a very dry site with a median PWV of ~2.4 mm, offering excellent conditions for IR observations when real-time PWV information is taken into account. Its pronounced seasonal variations are well-documented[1]. Similarly, the likelihood of clouds passing over Paranal is seasonally dependent.

*fkerber@eso.org; phone +49 89 32006757; fax +49 89 32006838; www.eso.org

Recently (July 5, 2012), a "dry episode" with PWV ~0.1 mm lasting for 12 hours has been reported by Kerber et al[3]. This was related to the excursion of Antarctic air to the Atacama Desert. During these spectacular conditions the transparency of the atmosphere will increase even outside the established atmospheric windows. One important aspect in such conditions is the homogeneity of PWV across the sky. For Paranal this has been studied for the first time in a contribution at this conference[3].

### 1.1 Low Humidity and Temperature Profiling microwave radiometer (LHATPRO)

The Low Humidity And Temperature PROfiling microwave radiometer (LHATPRO), manufactured by Radiometer Physics GmbH (RPG), measures the atmosphere at two frequency ranges focusing on two prominent emission features: a strong $H_2O$ line (183 GHz) and an $O_2$ band (51-58 GHz). Using 6 and 7 channels, respectively, the radiometer retrieves the profile of humidity and temperature up to an altitude of ~12 km[4,5]. The spatial resolution is given by the size of the radiometer beam (1.4° FWHM). For a full calibration with absolute standards an additional external calibration target, cooled down to the boiling point of liquid nitrogen (LN2), is used. The LHATPRO offers a measurement duty cycle of >97%. Details of the radiometer are described in Rose et al[5]. Before deployment of the LHATPRO, efforts to monitor water vapour on Paranal used dedicated standard star observations at the 8-m unit telescopes (UTs) with the spectroscopic instruments UVES, X-Shooter, CRIRES and VISIR. See ESO's instrumentation webpage: http://www.eso.org/sci/facilities/paranal/instruments.html for details. The method for deriving PWV from such spectra taken at various wavelengths by fitting the observed spectrum with an atmospheric radiative transfer model is described in Querel et al[6].

The LHATPRO water vapour radiometer (WVR) was commissioned during a 2.5 week period in October and November 2011 during which the calibration was tested with respect to balloon-borne radiosondes, the established standard in atmospheric physics[1]. In addition, ESO's LHATPRO is equipped with an IR radiometer (10.5 µm) to measure sky brightness temperature (IRT) down to -100° C. This allows for detection of high altitude clouds, e.g. cirrus, that consist of ice crystals but contain practically no liquid water. The IR radiometer can by fully synchronized for scanning with the microwave radiometer.

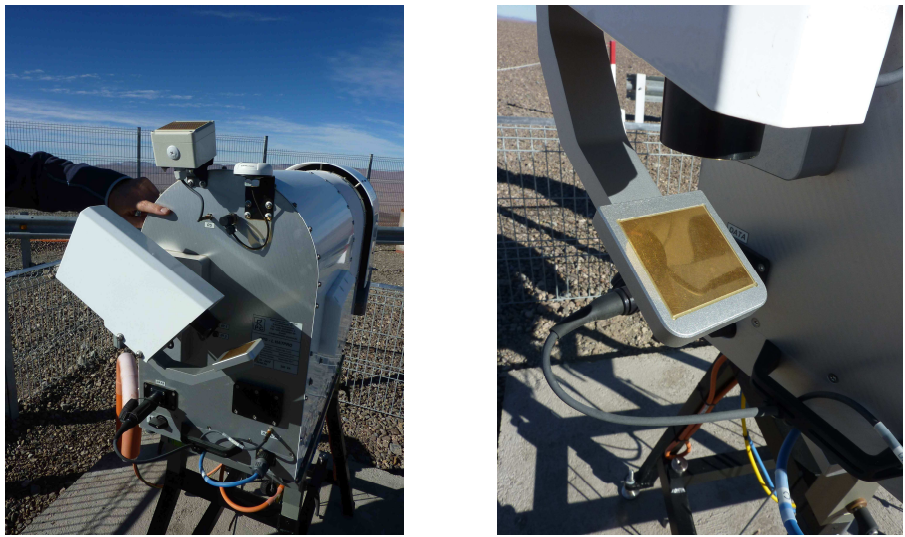

Figure 1. LHATPRO on Paranal (left), the white unit in the foreground is the IR camera. It can move in elevation in lock with the radiometer. Hence it can point to arbitrary positions together with the radiometer and perform sky scans. Close-up of the IR camera (right). In the center is the upward facing gold relay mirror. With this set-up the camera lens will always face downwards and is better protected from the elements.

The LHATPRO has an all-sky pointing capability that can be used to conduct user-defined scans of the sky in an automated fashion. The WVR can be commanded to repeat an observation sequence n-times resulting in months of unsupervised operations if needed. The IR channel is mounted on the same azimuth stage and can be instructed to perform its elevation scan in lock with the water vapour radiometer resulting in fully synchronous and parallel observations in the IR channel as well. The operational scheme currently set-up on Paranal and covering a period of 24 hours is the following:

- 2-D all-sky scan: step size of 12° azimuth, 12.5° elevation (duration 6 min, repeated every 6 h; 1.7% of day)
- Cone scan at 30° elevation (Hovmöller): step size 6° (duration 2.5 min, repeated every 15 min; 16.7% of day)
- Routine observations set to zenith-staring mode (81.6% of day)

For the current analysis we have used only zenith-staring observations but all-sky scans and Hovmöller scans can be used for similar analysis in the future. In particular the Hovmöller taken at higher air mass will be of interest.

### 1.2 Science operations and user-supplied atmospheric constraints

ESO uses so-called calibration plans that ensure that certain levels of precision are provided by service mode observations, e.g. 5% precision for photometry. If more demanding levels of accuracy are required for the scientific objective the astronomical user has to include dedicated calibration observations such as standard stars for photometry in his or her proposal. These observations are counted as observing time whereas the routine calibration observations required to meet the requirements of the calibration plan are absorbed by the observatory.

ESO has been successfully using the concept of service mode observations for many years now. There are two elements to it: a) users can provide observing constraints required to meet their scientific goals; and, b) a queue-based operational model allows the support astronomers at the observatory to flexibly react to changes in ambient conditions in order to optimally satisfy all requested constraints. In terms of clouds, ESO uses a classification scheme that allows users to ask for photometric conditions if quantitative measurements require it or to accept different levels of cloudiness if their science is not negatively affected by this.

### 1.3 ESO sky quality classification

In service mode the astronomer may specify a number of constraints (atmospheric seeing, lunar illumination, air mass, etc.) required for ensuring the success of the observations and their scientific value. In terms of clouds the following options are defined:

- Photometric:

  No visible clouds; transparency variations under 2%.
- Clear:

  Less than 10% of the sky above 2 air masses covered by clouds; transparency variations less than 10%.
- Variable, thin cirrus:

  Transparency variations between 10% and 20%.
- Variable, thick cirrus:

  Large transparency variations possible, equivalent to no constraint on the transparency conditions.

The selected condition has to be fulfilled for the duration of the observation, which is organized at ESO in observation blocks (OB) with a maximum length of 1 h, and not for a full night. Note that the definition of "Clear" does not require a cloudless sky.

The most demanding category is of course "Photometric" which is a requirement for quantitative photometry and usually requires the transmission to be stable to within a few percent for the whole night at any observatory. Hence the establishment of photometric conditions requires several steps. The photometric zero-point (ZP) is determined by observing established standard stars and comparing the throughput to a reference value – agreement better than 5% is required. On Paranal ZPs can be measured by any of the following optical and IR instruments: FORS2, VIMOS, ISAAC, NACO, HAWK-I, VIRCAM or OmegaCam. For further details: http://www.eso.org/sci/facilities/paranal/instruments.html

The measurement sequence for ZPs is as follows:

- a first measurement at low air mass determines if the throughput at low air mass satisfies the throughput constraint;
- if the throughput constraint is satisfied, a second measurement at high-air mass is the absolute minimum to estimate the extinction coefficient;
- a third measurement after the execution of the scientific OB requiring photometric conditions checks the stability of the atmospheric transparency.

In addition, a scientific OB requiring photometric conditions must be bracketed by standard star observations within 3 hours before and after the execution of the OB. In practice, instrumental effects will also be important in affecting the accuracy of photometry. The FORS accurate photometry project (FAP) has carefully analysed the possible systematic effects and demonstrates that 1-2% accuracy can be achieved with this instrument across the whole field of more than 6 arcmin[7].

From the above it is evident that it involves significant effort to ensure that photometric conditions are reliably determined. Specifically, it requires the observations of standards stars with large telescopes. Hence it will be attractive to have a stand-alone instrument that can help with assessing sky conditions in terms of clouds and provide continuous monitoring. The LHATPRO IR channel has some capabilities towards this end but we have just started to explore the automated analysis of sky brightness measurements for this purpose.

## 2. DETRENDED FLUCTUATION ANALYSIS

The detrended fluctuation analysis (DFA) method, first described by Peng et al.[8], is a powerful tool that can readily find otherwise hidden features in what, at first glance, might be assumed to be a random, irregular time series. This "self-affinity" in a data set can range from short intervals through to long-term effects. The approach was first used in biological applications related to characterizing seemingly pattern-less, DNA coding sequences. The method has gone on to be used in everything from financial systems to the analysis of atmospheric data sets. The method works as follows: a data sequence $y(t)$, of length $N$, is subdivided into pieces of equal length $\tau$. Each sub-unit of data is fit with a line $z(t) = at + b$ and then has that local trend subtracted from itself, detrending each segment.

This is applied to each sub-unit, across the entire sequence $y(t)$. This process is then fully repeated for a wide range of $\tau$ values. The mean square difference over each interval can be computed as:

$$F^2(\tau) = \frac{1}{\tau} \sum_{t=k\tau+1}^{(k+1)\tau} \{y(t) - z(t)\}^2$$

$$k = 0, 1, 2, \ldots, \left(\frac{N}{\tau} - 1\right)$$

where $k$ is the number of pieces in each sequence for a given $\tau$. The $F^2(\tau)$'s are averaged over all segments in each sequence. This results in the DFA function $\langle F^2(\tau) \rangle$ that represents fluctuation correlations as a function of the time interval $\tau$. The DFA function is expected to follow a power law[8]:

$$\langle F^2(\tau) \rangle^{1/2} \sim \tau^\alpha .$$

When the above relationship is plotted on a log-log graph, the $\alpha$-exponent is directly accessible as the slope of the resulting curve.

An $\alpha < 0.5$ implies anti-persistence in the signal, meaning an increase is likely to be followed by a decrease and vice versa. An $\alpha > 0.5$ implies persistence, meaning an increase is likely to be followed by a further increase in the signal and vice versa. At $\alpha = 0.5$, the next step, increase or decrease, is equally probable. For Gaussian white noise, $\alpha = 0$.

The behaviour of the $\alpha$-exponent as it relates to atmospheric conditions and its usage in identifying the presence of clouds and potentially classifying them has been previously demonstrated[9,10,11].

The α-values (slopes) for each DFA data set in the analysis that follows has been determined from the best-fit line over the τ-interval from 60–300 seconds. This fit range is depicted as the horizontal bar in the right panels of Figures 2–9.

# 3. RESULTS

For illustration we provide examples of different night conditions (Figs. 2–9) over Paranal. This is followed by a statistical analysis.

## 3.1 Thick clouds

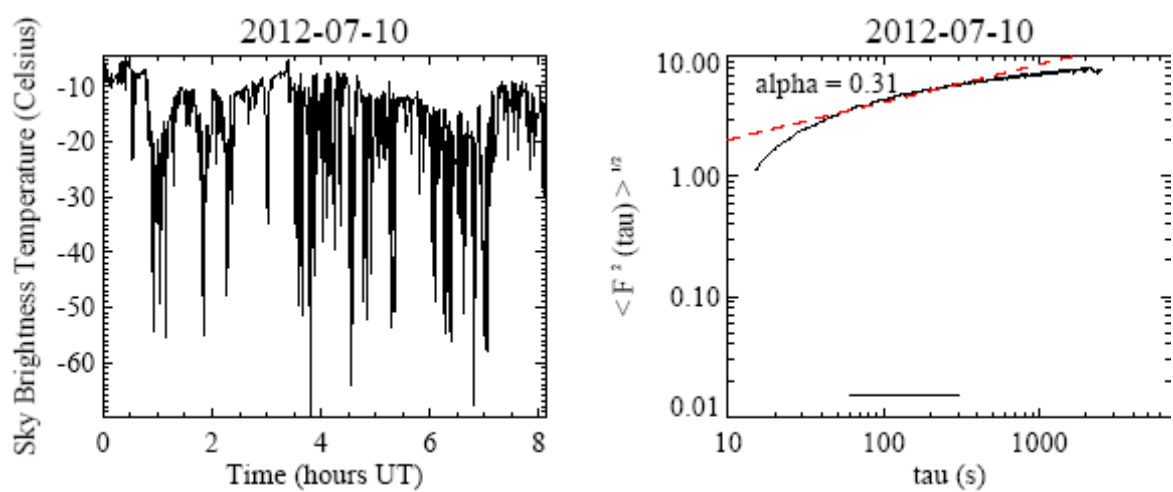

Figure 2. Example of thick conditions from July 10, 2012. The variability is high and hence the alpha parameter is relatively high at 0.31. The fit-range of 60 to 300 seconds is plotted as the horizontal bar in the right panel.

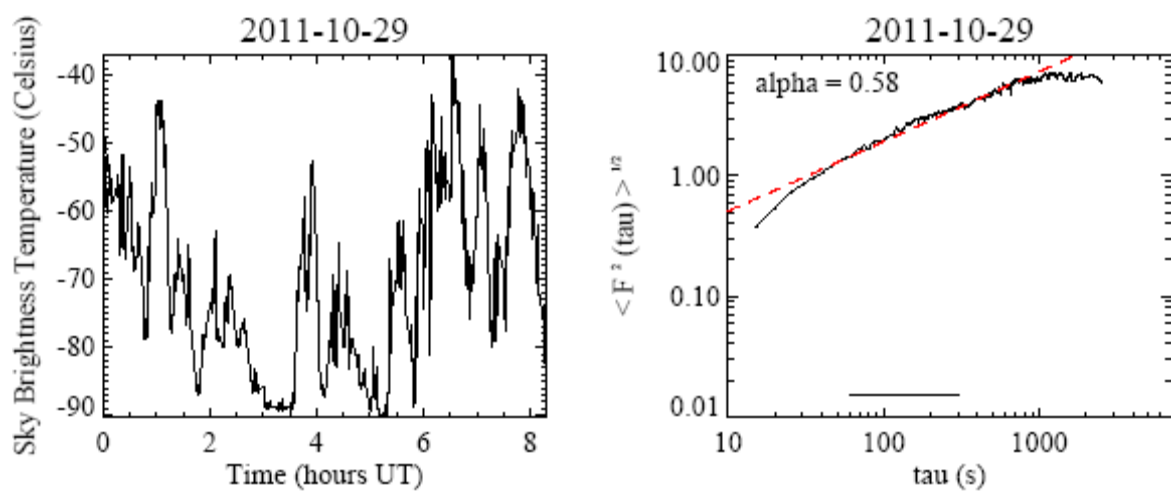

Figure 3. Example of thick conditions from October 29, 2011. The variability is high and the alpha parameter is high at 0.58.

## 3.2 Thin clouds

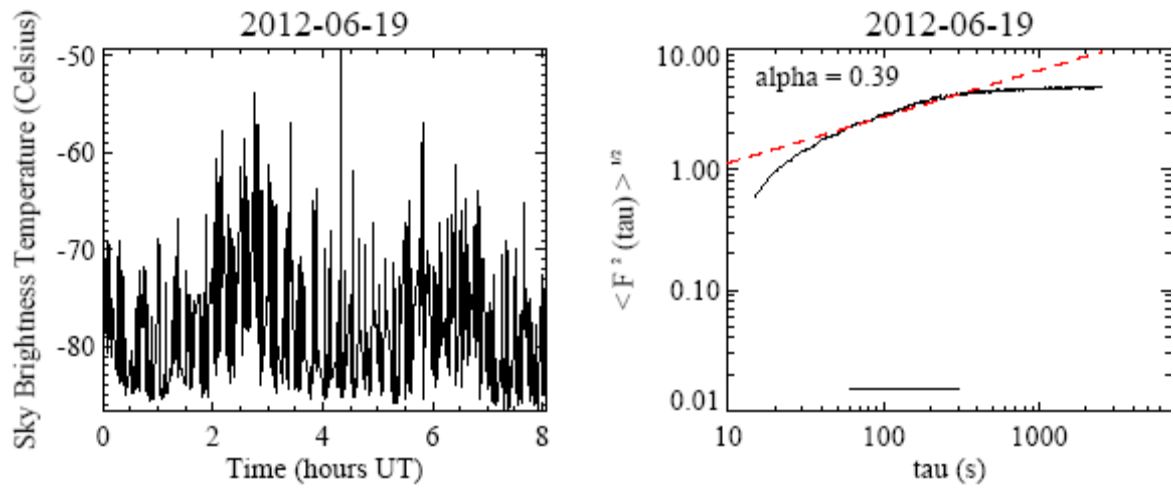

Figure 4. June 19, 2012 is an example of thin cloud conditions. The overall alpha for this 8 hour period is 0.39.

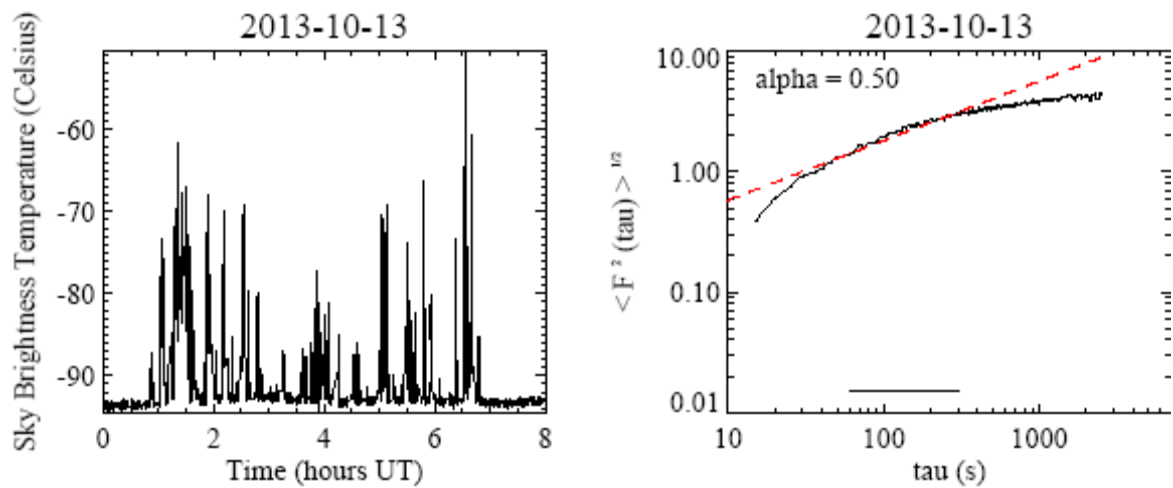

Figure 5. Another example of thin cloud conditions is October 13, 2013. The alpha in this case is 0.50.

## 3.3 Clear skies

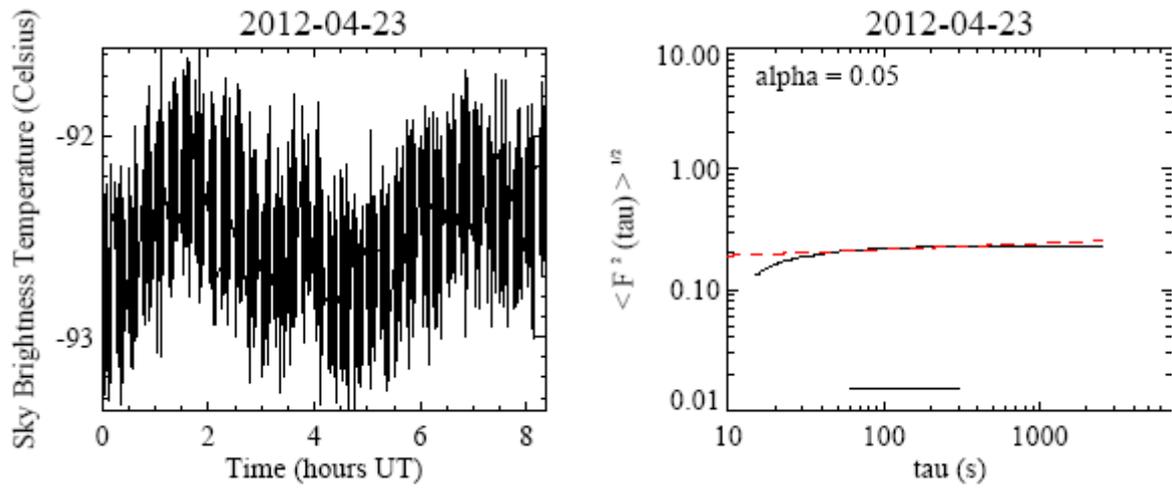

Figure 6. Example of clear sky conditions from April 23, 2012. The IRT variability is low (within 1 °C) and the DFA derived alpha parameter is 0.05.

## 3.4 Variable conditions

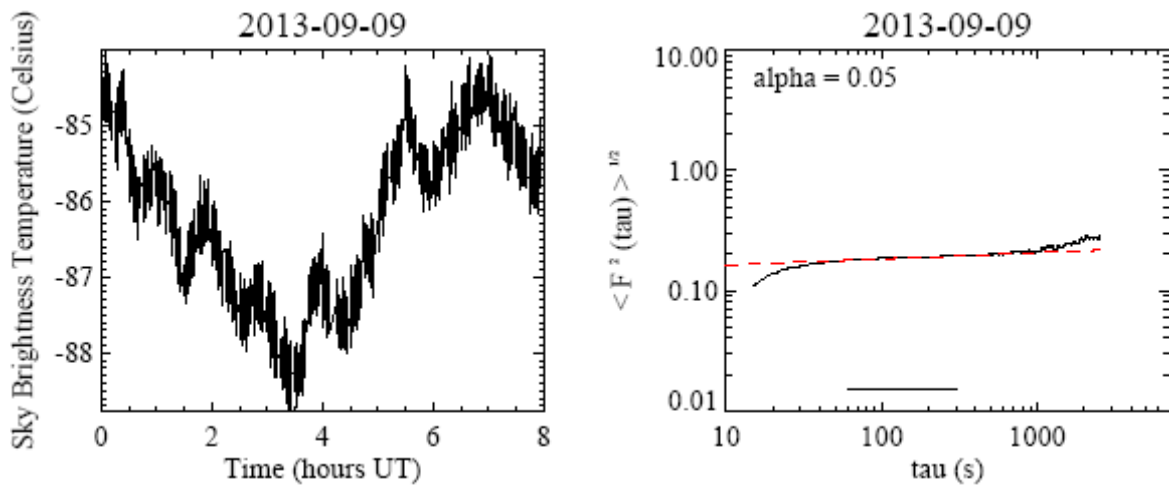

Figure 7. An example of IRT with a seemingly high variability which also results in a low alpha parameter value (0.05). In this figure the IRT from the night of September 9, 2013 varies by nearly 5 °C and shows a pronounced feature that has negligible effect on the calculated alpha value.

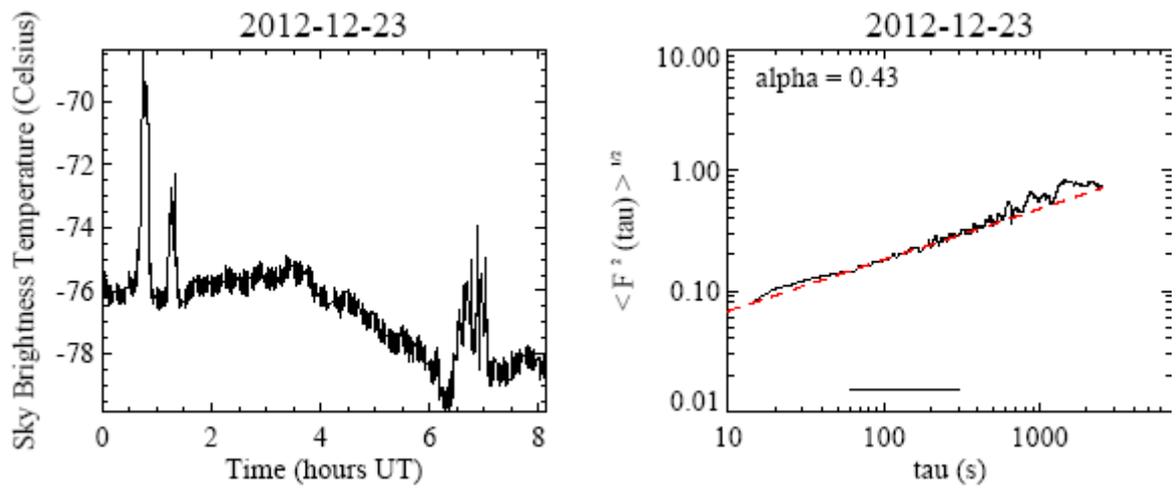

Figure 8. Example of clear skies interrupted by episodes of thin clouds on Dec 23, 2012. IRT time series (left) and DFA plot (right); the resulting alpha parameter is 0.43. This demonstrates that automated DFA analysis of full nights will also indicate episode of clouds by reporting an alpha significantly larger than for clear skies.

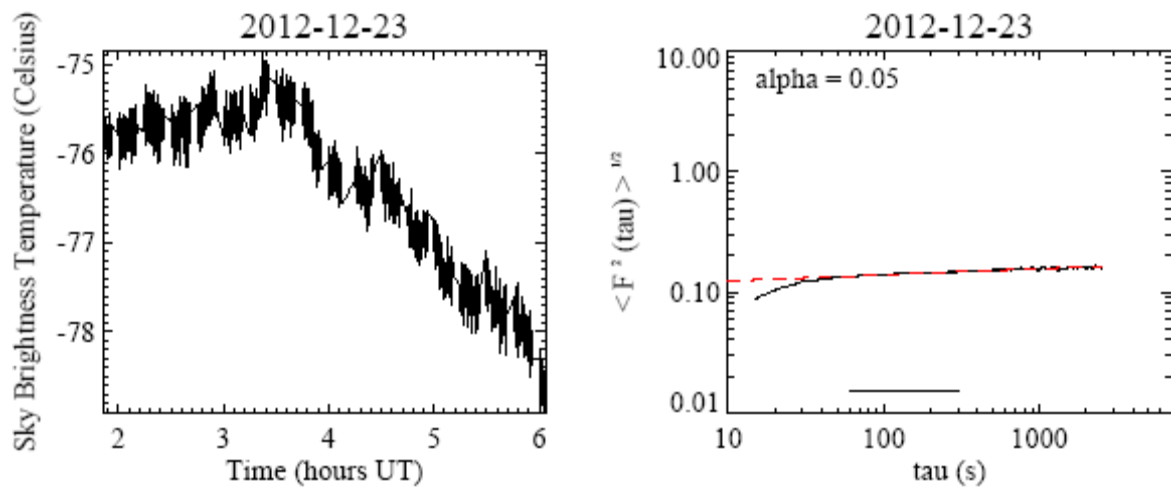

Figure 9. Close-up of cloudless period on Dec 23, 2012, the same night shown in Fig 8. IRT time series (left) and DFA plot (right); the resulting alpha parameter from this subset is 0.05, typical of cloudless conditions.

One important result is that DFA is robust with respect to variations in the IRT resulting from effects other than clouds, e.g. variations in PWV, see Fig 7. From the current analysis is seems unlikely that spurious detections of clouds can be triggered by other IRT variations.

Another interesting result regarding variability is shown in Figures 8 and 9. The DFA derived from an entire night (Fig.8) and from a subset of that night (Fig.9) are both plotted. When the entire night of data (8 hours) are used the resulting alpha is 0.43, representing cloudy and variable, while the clear subset (4 hours) gives an alpha of 0.05, denoting very clear conditions.

### 3.5 Statistical analysis of sky conditions over Paranal

For an observatory and its users it is very important to know what atmospheric conditions can be expected for the scientific observations. For service mode users this is particularly important since they can specify observing constraints on atmospheric conditions in advance.

It is well-known that Paranal offers a very large fraction of clear and photometric nights per year. Hence it is no surprise that we find a very large fraction of nights with $\alpha < 0.10$ indicating cloudless conditions (Fig. 10, Table 1).

The DFA gives a result that describes the behavior of a time-series. For the analysis to function there needs to be a relatively large amount of data to process, i.e. it is not possible to use 5 minutes of data and characterize the sky. However, for example, by including the recent three hours of past data that time series can be lengthened sufficiently to enable determination of a relevant alpha-parameter. Then, by continuing this analysis with a moving 3-hour window, and comparing the resulting alpha values, one could attempt to follow the evolution of variations across the sky. This is depicted in Figure 6 of Brocard et al.[10]. In such a scenario, IRT data could be used as a proxy for photometric zero-points reducing the need for standard star observations. Further analysis addressing the correlation between IRT and actual transmission are needed. This will be done using archival data on the photometric ZPs and the night logs recording sky conditions.

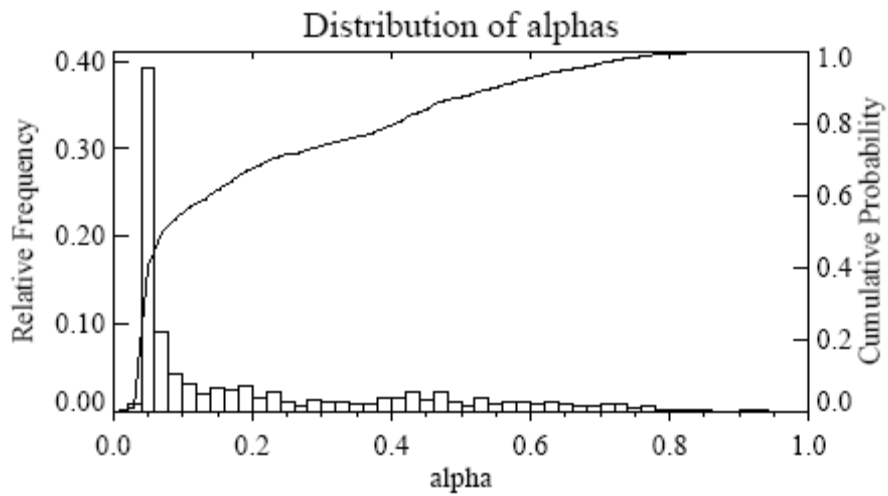

Figure 10. Histogram and cumulative distribution showing all alpha values for 880 nights (00:00 to 08:00 UT). A peak at 0.05 showing nearly 40% of nights on Paranal as being extremely clear conditions for the whole night.

Table 1. Percentiles of the distribution across the alpha values.

| Percentiles | alpha |
|---|---|
| 10 | 0.045 |
| 25 | 0.050 |
| **50** | **0.083** |
| 75 | 0.338 |
| 90 | 0.561 |

## 4. OUTLOOK

The DFA of LHATPRO IR measurements has demonstrated its ability to detect the presence of clouds of all kinds including very thin cirrus that are very difficult to detect otherwise, in particular when the moon is absent from the night sky. Further in-depth analysis is planned in order to use DFA for classifying clouds and their impact on atmospheric transmission. The alpha parameter is an excellent indication of time dependent variation in the sky brightness signal but is not directly probing the actual transmission of the cloud which is a physical properties. The IR channel of the LHATPRO also measures the temperature of the clouds which is related to the thickness and altitude of the clouds. We will use this additional information to extend our analysis. At this point we have only used the zenith staring observations. In determining whether a night is photometric (see discussion is section 1.3) it will be useful to also include the 2-D scans and the Hovmöller scans because those measurements contain information on spatial variation and stability.

On the technical side, RPG is investigating the possibility of expanding the capabilities of the IR channel by adding a second filter band to better discriminate between different kinds of clouds.

The ultimate accuracy in terms of photometry can be achieved by measuring the actual atmospheric transmission along the line-of-sight towards the target. The LHATPRO WVR is capable of all-sky pointing and tracking and thus line-of-sight support. The spatial resolution is limited by the size of the radiometer beam (1.4° FWHM). This mode has been technically implemented by RPG and has been tested on Paranal. Full implementation of such a mode will require careful planning and would have to be justified by clear scientific gain.

## 5. SUMMARY

We have used IR data from the camera on the LHATPRO to explore quantitative means of cloud detection and classification. The alpha parameter produced by detrended fluctuation analysis has been shown to be a sensitive discriminator for all kinds of clouds including very thin cirrus – the most common form of clouds over Paranal. For cloudless skies we find $\alpha \leq 0.10$. A classification of clouds based on alpha will require a more in-depth analysis and the use of other available parameters, for instance the IRT directly.

**Acknowledgements:** RQ acknowledges funding from Conicyt through Fondecyt grant 3120150.

endorsement by the European Southern Observatory, nor is it implied that the materials or equipment identified are necessarily the best available for the purpose.